\documentstyle[12pt,epsf]{article}
\newcommand{\lsim}{\,{\buildrel < \over {_\sim}}\,}
\newcommand{\gsim}{\,{\buildrel > \over {_\sim}}\,}

\begin{document}

\begin{titlepage}
\begin{flushright}
CERN-TH/97-345\\
JYFL-2/98\\
hep-ph/9802350\\
\end{flushright}
\begin{centering}
\vfill

{\bf SCALE EVOLUTION OF NUCLEAR PARTON DISTRIBUTIONS}

\vspace{0.5cm}
K.J. Eskola$^{\rm a,b,}$\footnote{kjeskola@ux.phys.jyu.fi},
V.J. Kolhinen$^{\rm a,}$\footnote{kolhinen@ux.phys.jyu.fi} and
P.V. Ruuskanen$^{\rm a,}$\footnote{ruuskanen@jyfl.jyu.fi}

\vspace{1cm}
{\em $^{\rm a}$ Department of Physics, University of Jyv\"askyl\"a,\\
P.O.Box 35, FIN-40351 Jyv\"askyl\"a, Finland\\}

\vspace{0.3cm}

{\em $^{\rm b}$ CERN/TH, CH-1211 Geneve 23, Switzerland\\}

\vspace{1cm}
{\bf Abstract}

\end{centering}
Using the NMC and E665 nuclear structure function ratios $F_2^A/F_2^D$
and $F_2^A/F_2^{\rm C}$ from deep inelastic lepton-nucleus collisions,
and the E772 Drell--Yan dilepton cross sections from proton-nucleus
collisions, and incorporating baryon number and momentum sum rules, 
we determine nuclear parton distributions at an initial scale $Q_0^2$.
With these distributions, we study QCD scale evolution of nuclear parton
densities. The emphasis is on small values of $x$, especially
on scale dependence of nuclear shadowing. As the main result,
we show that a consistent picture can be obtained within the leading 
twist DGLAP evolution, and in particular, that the calculated
$Q^2$ dependence of $F_2^{\rm Sn}/F_2^{\rm C}$ agrees very well 
with the recent NMC data.

\vspace{0.3cm}\noindent

\vfill
\noindent
CERN-TH/97-345\\
JYFL-2/98\\
February 1998

\end{titlepage}

\section{Introduction}

Structure function $F_2^A$ has been measured in deep inelastic
lepton-nucleus scatterings for a wide range of nuclei $A$
\cite{ARNEODO94}. The ratios of $F_2^A$ to the structure function of
deuterium, $F_2^A(x,Q^2)/F_2^D(x,Q^2)$,  reveal clear deviations
from unity. This indicates that parton distributions of bound
nucleons are different from the ones of free nucleons:
$xf_{i/A}(x,Q^2) \neq xf_{i/p}(x,Q^2)$. The nuclear effects
in the ratio $F_2^A/F_2^D$ are usually divided into the following
regions in Bjorken $x$:

\begin{itemize}
\item shadowing; a depletion  at $x \lsim 0.1$,
\item anti-shadowing; an excess at $0.1 \lsim x \lsim 0.3$,
\item EMC effect; a depletion at $0.3 \lsim x\lsim0.7$
\item Fermi motion; an excess towards $x\rightarrow1$ and beyond.
\end{itemize}

At the moment, there is no unique theoretical description of
these effects;  it is believed that different mechanisms are
responsible for them in different kinematic regions. For an
overview of the existing data and different  models, we refer
the reader to Ref.~\cite{ARNEODO94}.

While the $x$ dependence of the nuclear effects in $F_2^A$ was
observed already in the early measurements \cite{EARLY}, the $Q^2$
dependence is much weaker and has therefore been much more
difficult to detect. Only recently the first observation of 
a $Q^2$ dependence of the ratio $F_2^{\rm Sn}/F_2^{\rm C}$ has 
been  published by the New Muon Collaboration (NMC) \cite{NMC96}.
From the point of view of nuclear parton distributions,
a high-precision measurement of the scale dependence of $F_2^A/F_2^D$ or
 $F_2^A/F_2^{\rm C}$ is very important for pinning down the gluon
distributions in nuclei \cite{KJE}.  In Ref.~\cite{PIRNER}
preliminary NMC data  \cite{MUCKLICH} and the small-$x$ limit of 
the DGLAP equations \cite{DGLAP} were used to determine the gluon 
ratio $xg_{\rm Sn}/xg_{\rm C}$ from the measured  
$Q^2$-evolution of  $F_2^{\rm Sn}/F_2^{\rm C}$.

In this paper, our goal is to study whether the observed
$Q^2$ evolution  of $F_2^{\rm Sn}/F_2^{\rm C}$ \cite{NMC96} 
is consistent  with the leading twist  DGLAP-evolution.  
By using the data from deeply inelastic lepton-nucleus scattering 
\cite{NMCre}-\cite{E665} and from the Drell--Yan process in $pA$ 
collisions  \cite{E772}, and by  simultaneously requiring baryon 
number and momentum conservation \cite{KJE,FSL}, we will first 
determine a set of nuclear parton distributions at an initial scale 
$Q_0^2$. When doing so, we want to avoid using any specific model 
for the nuclear effects. We then evolve the initial distributions 
up to higher scales by using lowest order DGLAP equations without 
parton fusion corrections \cite{GLRMQ}. We will show explicitly 
how the nuclear effects in $F_2^A/F_2^D$ and the similarly defined 
ratio for the Drell--Yan cross sections evolve in $Q^2$, and make 
a comparison with the data. Our procedure leads to a consistent 
picture, and as the main result, we will show that a very good 
agreement  is obtained between our leading twist QCD approach 
and the NMC data  \cite{NMC96} for the $Q^2$ dependence of  
$F_2^{\rm Sn}/F_2^{\rm C}$.

Scale evolution of nuclear effects has been studied already earlier
\cite{QIU,FSL,KJE} but since then the parton distributions of {\em
proton} have become much better known, especially at small values 
of $x$, where a rapid rise of the structure function $F_2^p$ has been
observed at HERA \cite{HERA}. Consequently, also the gluon
distributions can be much better determined. So far, there has been 
no signs in the HERA data for the need of parton fusion corrections 
\cite{GLRMQ} to the evolution equations down to $Q^2\sim 1$~GeV$^2$ 
and $x\sim 10^{-4}$ \cite{HERAlowQ}. The situation has also improved 
regarding nuclear data; more high-precision  nuclear data for several 
mass numbers $A$ are available \cite{NMCre}-\cite{E665} at small
values of $x$ where our main interest is focussed. The QCD scale 
evolution of nuclear effects in parton distributions, 
especially with specific models for the initial distributions at 
$Q^2=Q_0^2$, has been studied in \cite{FSL,KUMANO,INDUMATHI}.
See also Refs.~\cite{LV}-\cite{RVMcL}.

\section{Nuclear parton distributions at $Q_0^2$}

As the first task, we determine the nuclear parton distributions 
at an initial scale $Q_0^2$  as model-independently as possible. 
It turns out that some assumptions are needed about the nuclear effects
on the initial  distributions of individual parton flavours; we
will try to be quite explicit on these assumptions.

\begin{figure}[b]
\vspace{-4cm}
\centerline{\hspace*{0cm} \epsfxsize=16cm\epsfbox{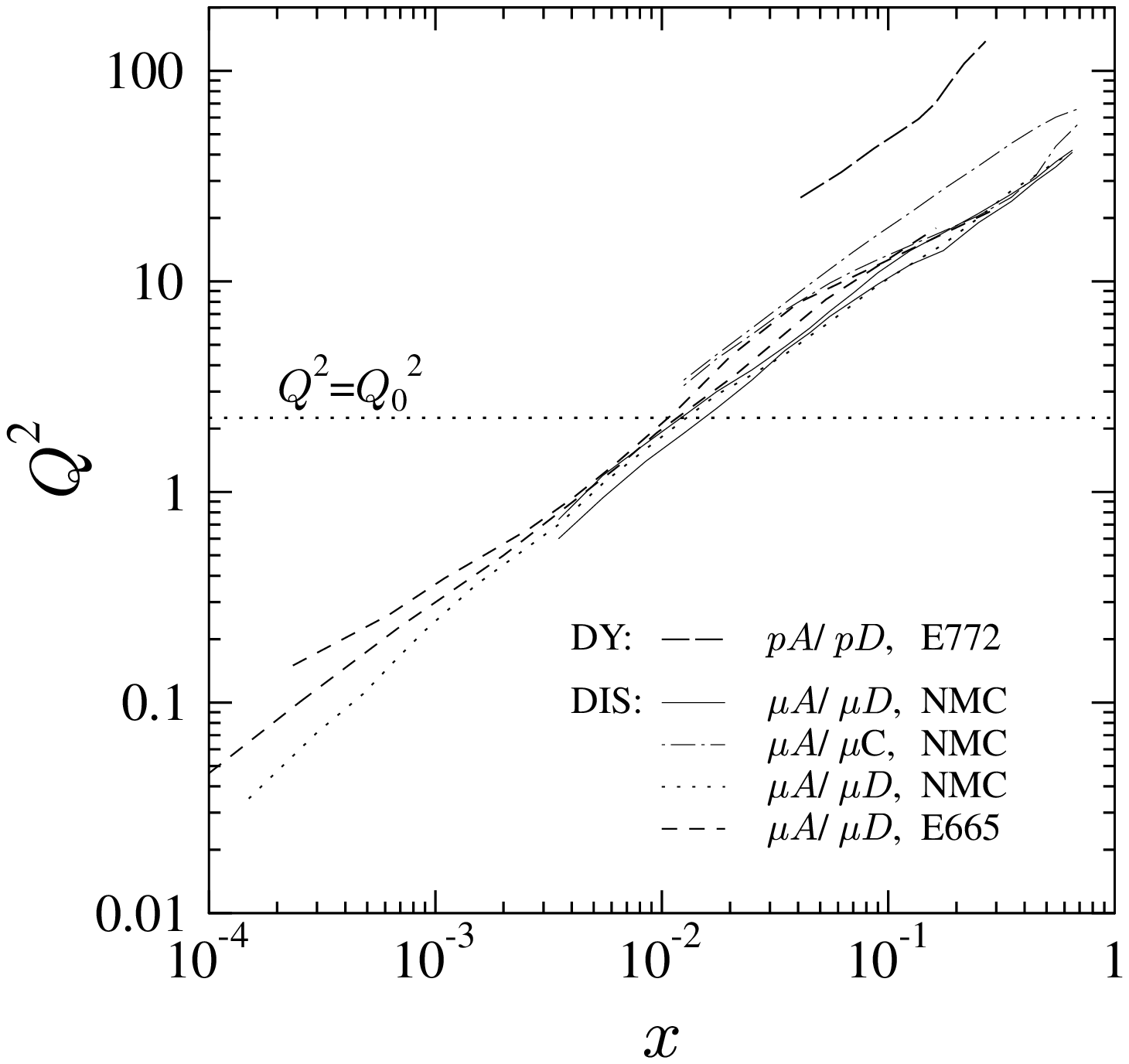}}
\vspace{-3cm}
\caption[a]
{{\small 
Typical correlation of the scale $\langle Q^2\rangle$ and $x$ in 
measurements of $F_2^A(x,Q^2)$ in  deeply inelastic $lA$ scatterings 
and  correlation of the invariant mass $\langle Q^2\rangle$ and $x=x_2$
of Drell-Yan cross sections measured in $pA$ collisions. 
The correlations in some of the NMC data \cite{NMCre} (solid lines), 
\cite{NMCsys} (dotted-dashed), \cite{NMCsat} (dotted) and in some of the 
E665 data  \cite{E665sat} (dashed), and in the E772 data \cite{E772,PMcG} 
(long dashed) are shown.
The horizontal dotted line illustrates the initial scale $Q_0^2$ we 
have chosen and above which we perform the DGLAP evolution of nuclear 
parton densities.
}}
\label{DATA}
\end{figure}

Fig.~\ref{DATA} illustrates the problem: both the deep inelastic 
lepton-nucleus data (DIS) and the proton-nucleus Drell--Yan
data (DY) lie on some curves in the $(x,Q^2)$-plane, determined 
by the kinematical limits and the experimental acceptances.  
In particular, they are not along a constant $Q^2$ line, as would 
be preferable for solving the DGLAP evolution.  Therefore, the 
initial distributions have to be determined iteratively
in such a way that the scale evolved distributions are
consistent with the data. Like in the case of a free proton, the
initial parton distributions at the chosen $Q_0^2$ serve as
nonperturbative input.  Our aim here is mainly to see whether a
consistent description based on leading twist QCD scale evolution can
be obtained.  Therefore we do not attempt to implement a
$\chi^2$-minimization procedure in determining the initial 
distributions but plan to return to this in the future.

The ratio of the structure function $F_2^A$ per nucleon in a 
nucleus $A$ with $Z$ protons and $A-Z$ neutrons, and $F_2^D$ 
of deuterium can be written as
\begin{eqnarray}
R_{F_2}^A(x,Q^2)
&\equiv& \frac{F_2^A(x,Q^2)}{F_2^D(x,Q^2)}\\
&=&\frac{[F_2^{p/A}(x,Q^2) + F_2^{n/A}(x,Q^2)]
+({2Z}/{A}-1)[F_2^{p/A}(x,Q^2) - F_2^{n/A}(x,Q^2)]}
      {F_2^{p/D}(x,Q^2) + F_2^{n/D}(x,Q^2)}.
\end{eqnarray}
In the lowest order in QCD-improved parton model
$F_2(x,Q^2) = \sum_q e_q^2 [xq(x,Q^2) + x\bar q(x,Q^2)]$.
Below the charm-mass threshold we then have
\begin{equation}
R_{F_2}^A(x,Q^2) = \frac{5(u_A+\bar u_A + d_A +\bar d_A) + 4s_A
+(\frac{2Z}{A}-1)3(u_A+\bar u_A - d_A-\bar d_A)}
     {5(u+\bar u + d +\bar d) + 4s},
\label{RF2}
\end{equation}
where  $u\equiv u(x,Q^2)$ is the known distribution of $u$ quarks
in a free proton, and $u_A\equiv u_{p/A}(x,Q^2)$ is the average
$u$-quark distribution in a bound proton of a nucleus $A$, and 
similarly for other quark flavours. For isoscalar nuclei 
$d_{n/A}= u_A$ and $u_{n/A}= d_A$, and in the formula (\ref{RF2}) 
above we have assumed that this is a good approximation for the
non-isoscalar nuclei as well. Nuclear (shadowing) effects in 
deuterium have been  neglected in Eq.~(\ref{RF2}). These have 
been studied in \cite{BK}, where it is shown that the shadowing 
corrections to $(F_2^p+F_2^n)/2$ are of order 1~\% at $x\gsim 0.007$. 
Cumulative effects for $x > 1$ will also be neglected as well 
as the binding energy effects when using $M_A/A\approx m_p\approx m_n$.

We define the nuclear valence quark distributions in a usual way,
$q_V^A \equiv q_A-\bar q_A$, assuming that any differences between
quarks and antiquarks in the nuclear sea \cite{BRODSKY}
can be neglected.  Let us also  define the following
ratios for each sea- and valence-quark flavour:
\begin{eqnarray}
R_{\bar q}^A (x,Q^2) &\equiv& \frac{\bar q_A(x,Q^2)}{\bar q(x,Q^2)}, 
\label{RqA} 
\\
R_{q_V}^A (x,Q^2)    &\equiv& \frac{q_V^A(x,Q^2)}{q_V(x,Q^2)}.
\end{eqnarray}
For later use, it is convenient to define also the corresponding
ratios for the total nuclear valence-quark and light sea-quark
distributions:
\begin{eqnarray}
R_V^A(x,Q^2) &\equiv& 
\frac{u_V^A(x,Q^2)+d_V^A(x,Q^2)}{u_V(x,Q^2)+d_V(x,Q^2)},\\
R_S^A(x,Q^2) &\equiv& 
\frac{\bar u_A(x,Q^2) + \bar d_A(x,Q^2) +\bar s_A(x,Q^2)}
{\bar u(x,Q^2) + \bar d(x,Q^2) +\bar s(x,Q^2)},
\label{RVA}
\end{eqnarray}
and similarly for the sum $\bar u+\bar d$, and for the differences
(appearing in the non-isoscalar part) $\bar u-\bar d$ and $u_V-d_V$
\begin{eqnarray}
R_{\bar u+\bar d}^A(x,Q^2)
&\equiv&
\frac{\bar u_A(x,Q^2)+\bar d_A(x,Q^2)}{\bar u(x,Q^2)+\bar d(x,Q^2)}, \\
R_{\bar u-\bar d }^A(x,Q^2)
&\equiv&
\frac{\bar u_A(x,Q^2)-\bar d_A(x,Q^2)}{\bar u(x,Q^2)-\bar d(x,Q^2)}, \\
R_{u_V-d_V }^A(x,Q^2)
&\equiv&
\frac{u_V^A(x,Q^2)-d_V^A(x,Q^2)}{u_V(x,Q^2)-d_V(x,Q^2)}.
\label{RNIS}
\end{eqnarray}
The relations between $R_V^A,R_{u_V-d_V}^A$ and  $R_{u_V}^A, R_{d_V}^A$,
and between $R_{\bar u+\bar d}^A,R_{\bar u-\bar d}^A$ and 
$R_{\bar u}^A,R_{\bar d}^A$ are obvious.  Then Eq.~(\ref{RF2}) becomes
\begin{eqnarray}
R_{F_2}^A(x,Q^2) &=& A_V^{IS}(x,Q^2) R_V^A(x,Q^2) +
A_{ud}^{IS}(x,Q^2) R_{\bar u+\bar d}^A(x,Q^2)+ A_s(x,Q^2)  R_s^A(x,Q^2)
\nonumber\\
&&+(\frac{2Z}{A}-1) [A_V^{NIS}(x,Q^2) R_{u_V-d_V }^A(x,Q^2) +
A_{ud}^{NIS}(x,Q^2) R_{\bar u-\bar d}^A(x,Q^2)],
\label{RF2A}
\end{eqnarray}
where the coefficients are known: 
\begin{eqnarray}
A_V^{IS}(x,Q^2) &=& 5[u_V(x,Q^2)+d_V(x,Q^2)]/N_{F_2}(x,Q^2)\\
A_{ud}^{IS}(x,Q^2) &=& 10[\bar u(x,Q^2)+\bar
d(x,Q^2)]/N_{F_2}(x,Q^2)\\
A_s(x,Q^2) &=& 4s(x,Q^2)/N_{F_2}(x,Q^2)\\
A_V^{NIS}(x,Q^2) &=& 3[u_V(x,Q^2)-d_V(x,Q^2)]/N_{F_2}(x,Q^2)\\
A_{ud}^{NIS}(x,Q^2) &=& 6[\bar u(x,Q^2)-\bar d(x,Q^2)]/N_{F_2}(x,Q^2)\\
N_{F_2}(x,Q^2)&=& 5[u_V(x,Q^2) + d_V(x,Q^2)] +10[\bar u(x,Q^2)+\bar d(x,Q^2)]
             + 4s(x,Q^2).
\end{eqnarray}

In order to have further constraints for the individual nuclear
ratios above, we will consider differential Drell--Yan cross sections in
$pA$ collisions. For a nucleus with $Z$ protons and $A-Z$ neutrons
the ratio of cross section to that for deuterium can be written in the
lowest order as\footnote{In Ref.~\cite{KJE}, the corresponding
formula is incorrect,
but it has the correct large-$x_F$ limit, which was used only.}
\begin{eqnarray}
&\!\!\!\!\!R&\!\!\!\!\!_{DY}^A(x_2,Q^2) \equiv  \frac
{\frac{1}{A} {d\sigma^{pA}_{DY}}/{dx_2dQ^2}}
{\frac{1}{2} {d\sigma^{pD}_{DY}}/{dx_2dQ^2}}
\nonumber\\
&&\!\!\!\!\!\!\!
 = \{4[u_1(\bar u_2^A+\bar d_2^A)+ \bar u_1(u_2^A+d_2^A)] +
  [d_1(\bar d_2^A+\bar u_2^A) + \bar d_1(d_2^A+u_2^A)] +
4s_1s_2^A +...\}/N_{DY}  \nonumber\\
&&\!\!\!\!\!\!\!+ (\frac{2Z}{A}-1)
\{4[u_1(\bar u_2^A-\bar d_2^A) + \bar u_1(u_2^A-d_2^A)]+
[d_1(\bar d_2^A-\bar u_2^A) +  \bar d_1(d_2^A-u_2^A)]\}/N_{DY}
\end{eqnarray}
where
\begin{equation}
N_{DY} = 4[u_1(\bar u_2+\bar d_2)+\bar u_1(u_2+d_2)] +
  [d_1(\bar d_2+\bar u_2) +\bar d_1(d_2+u_2)] + 4s_1s_2 + ...
\end{equation}
and we have used the notation $q_i^{(A)}\equiv q_{(A)}(x_i,Q^2)$
for $i = 1,2$ and
$q=u,d,s,...$. The variable $Q^2$ is the invariant mass of the lepton pair.
The target (projectile) momentum fraction is $x_2\ (x_1)$, and
$x_1 = Q^2/(s x_2)$.
With the definitions in Eqs.~(\ref{RqA})-(\ref{RNIS}), we obtain
\begin{eqnarray}
R_{DY}^A(x,Q^2)& = &
B_{ud}^{IS}(x_1,x_2,Q^2) R_{\bar u+\bar d}^A(x_2,Q^2)+
B_{V}^{IS}(x_1,x_2,Q^2) R_{V}^A(x_2,Q^2)+
\nonumber\\
&& + B_{s}(x_1,x_2,Q^2) R_{s}^A(x_2,Q^2)+
(\frac{2Z}{A}-1) [B_{ud}^{NIS}(x_1,x_2,Q^2) R_{\bar u-\bar d}^A(x_2,Q^2)+
\nonumber\\
&& + B_{V}^{NIS}(x_1,x_2,Q^2) R_{u_V-d_V}^A(x_2,Q^2)]
\label{RDY}
\end{eqnarray}
where the coefficients are:
\begin{eqnarray}
B_{ud}^{IS}(x_1,x_2,Q^2) & =  &
[4(u_1+\bar u_1)+d_1+\bar d_1](\bar u_2+\bar d_2)/N_{DY}\\
B_{V}^{IS}(x_1,x_2,Q^2) & =& (4\bar u_1+\bar d_1)(u_{V2}+d_{V2})/N_{DY}\\
B_{s}(x_1,x_2,Q^2) & = & 4s_1s_2/N_{DY}\\
B_{ud}^{NIS}(x_1,x_2,Q^2) & =  &
[4(u_1+\bar u_1)-d_1-\bar d_1](\bar u_2-\bar d_2)/N_{DY}\\
B_{V}^{NIS}(x_1,x_2,Q^2) & = & (4\bar u_1-\bar d_1)(u_{V2}-d_{V2})/N_{DY}\\
N_{DY}    & = &   [4(u_1+\bar u_1)+d_1+\bar d_1](\bar u_2+\bar d_2)
+ 4s_1s_2 + (4\bar u_1+\bar d_1)(u_{V2}+d_{V2})
\end{eqnarray}
with $q_{V2}\equiv q_V(x_2,Q^2)$ and
$q_1 = q_V(x_1,Q^2) + \bar q(x_1,Q^2)$, $q=u,d$.

The DIS data \cite{NMCre,NMCsys,SLACre,NMC96} which we use in
determining  the nuclear ratios, are approximately corrected 
for non-isoscalar effects, so we may consider
isoscalar nuclei first. In Eqs.~(\ref{RF2A}) and (\ref{RDY}) the terms
proportional to $(2Z/A-1)$  can be dropped, and we are left
with $R_V^A, R_{\bar u+\bar d}^A$ and $R_s^A$ only. Without
additional information we cannot fix three ratios from two
equations. Further constraints could in principle be obtained
from the measurements of nuclear structure functions 
$F_2^{\nu A}/F_2^{\nu D}$ and $F_3^{\nu A}/F_3^{\nu D}$
with neutrino beams \cite{BEBC}, but more statistics and mass number 
systematics would be needed. Instead, as a first approximation,
we take for the sea quarks 
$R_s^A(x,Q_0^2) = R_{\bar u+\bar d}^A(x,Q_0^2)$. For evolving
each flavour separately, we will also assume that initially
$R_{\bar q}^A(x,Q_0^2)=R_S^A(x,Q_0^2)$,
and similarly for the valence quarks
$R_{u_V}^A(x,Q_0^2) = R_{d_V}^A(x,Q_0^2) = R_V^A(x,Q_0^2)$.
It should be emphasized that this approximation is needed
{\em only} at the initial scale $Q_0^2$, in determining the
initial distributions for the DGLAP evolution. In this paper 
we use, for simplicity, the parton distribution set GRV-LO 
\cite{GRVLO}, where the massive quarks evolve as massless 
ones above each mass-threshold. This set also assumes 
$\bar u = \bar d$, so $R_{\bar u}^A = R_{\bar d}^A$ at 
all scales for this particular set.

In this approximation, we obtain two equations for the ratios at an
initial scale $Q_0^2$,
\begin{eqnarray}
\hspace{-0.5cm}R_{F_2}^A(x,Q^2_0) &=& A_V^{IS}(x,Q^2_0) R_V^A(x,Q^2_0) +
[A_{ud}^{IS}(x,Q^2_0) + A_s(x,Q^2_0)]  R_S^A(x,Q^2_0),
\label{RF2appro}
\\
\hspace{-0.5cm}R_{DY}^A(x,Q^2_0)& = &
B_{V}^{IS}(x_1,x,Q^2_0) R_{V}^A(x,Q^2_0) +
[B_{ud}^{IS}(x_1,x,Q^2_0) + B_{s}(x_1,x,Q^2_0)]R_S^A(x,Q^2_0)
\label{RDYappro}
\end{eqnarray}
where $Q_0^2$ is chosen below the charm threshold.
Eqs.~(\ref{RF2appro}) and (\ref{RDYappro})  would fix $R_V^A$ and $R_S^A$
if the DIS and  DY data would lie at the same values of $x$ and
$Q_0^2$. Since this is not the case (see Fig.~\ref{DATA}), 
the initial profiles $R_S^A(x,Q_0^2)$ and $R_V^A(x,Q_0^2)$ are
determined iteratively after comparing the evolved distributions with
the data. As is clear from Eqs.~({\ref{RF2appro}})
and (\ref{RDYappro}), the initial nuclear ratios acquire some
dependence  on the particular parton distribution set used for the
proton. With the modern sets, however, this dependence can be
expected to be quite weak.

As a further constraint for nuclear valence quarks \cite{FSL,KJE},
baryon number conservation 
\begin{equation}
\int_0^1 dx [u_V(x,Q_0^2) + d_V(x,Q_0^2)] R_V^A(x,Q_0^2)
= \int_0^1 dx [u_V(x,Q_0^2) + d_V(x,Q_0^2)]
= 3.
\label{CHARGE}
\end{equation}
will also be required \cite{FSL,KJE}.

In practice, our iteration procedure for determining the ratios
$R_{S}^A(x,Q_0^2)$ and $R_V^A(x,Q_0^2)$ proceeds with
the following steps:
\begin{itemize}
\item make an ansatz for $R_{F_2}^A(x,Q_0^2)$ based on the DIS data; 
\item decompose $R_{F_2}^A(x,Q_0^2)$ into $R_V^A$ and $R_S^A$, and
      constrain $R_V^A$ with baryon number conservation;
\item estimate $R_G^A(x,Q_0^2)$ (see the discussion below);
\item perform DGLAP evolution with the obtained initial 
      nuclear parton distributions;
\item constrain  $R_S^A(x,Q_0^2)$ and  $R_V^A(x,Q_0^2)$ 
      with the DY data.
\end{itemize}

First, we introduce an initial parametrization for $R_{F_2}^A(x,Q_0^2)$.
We choose $Q_0^2=2.25$~GeV$^2$, which is the charm-mass threshold
for the GRV-LO set. Rather than trying to  make a separate analysis 
for each nucleus, we parameterize also the $A$ dependence of $R_{F_2}^A$. 
The  functional form is given in the Appendix.  Note also, that as the 
existing parametrizations  \cite{SMIRNOV,VARY} are fits to the 
data along the kinematical curves in Fig.~\ref{DATA}, we cannot
use them directly for obtaining the distributions at fixed scale
$Q_0$. Furthermore, we have to make an assumption on the behaviour of
$R_{F_2}^A(x,Q_0^2)$ at $x\lsim 10^{-3}$. In this region a saturation 
of shadowing has been observed \cite{E665sat,NMCsat} at nonperturbative 
scales,  $\langle Q^2\rangle\ll Q_0^2$ (see Figs.~\ref{DATA} and 
\ref{RF2x}). Motivated by this, we will also assume  saturation of 
shadowing in $R_{F_2}^A(x,Q_0^2)$, keeping in mind, however, that 
consistency  of this assumption must be verified after evolving 
the distributions (see Figs.~\ref{RF2Q2}).

We then decompose $R_{F_2}^A$ into $R_V^A$ and $R_S^A$ according to
Eq.~(\ref{RF2appro}). First observation is that we cannot have 
$R_V^A=R_{F_2}^A$ at all values  of $x$, because up to 9~\% of 
baryon number would be missing for the heaviest nuclei. 
Second observation is that $R_{F_2}^A\approx R_V^A \, (R_S^A)$ at large 
(small) values of $x$. In practice, in the region of large (small) 
$x$, it becomes impossible to determine $R_S^A\, (R_V^A)$ directly 
from the data for $R_{F_2}^A$, because of the negligible sea 
(valence)-quark contribution. Therefore, we apply a piecewize 
construction: 
at $x<x_p\sim 0.1$ we fix $R_V^A$ with the same functional form as 
for $R_{F_2}^A$ but with different parameters. Then, through 
Eq.~(\ref{RF2appro}) $R_S^A$ becomes fixed. 
At  $x_p<x<x_{eq}\sim 0.4$ we fix $R_S^A$ in turn,
with a simple form of a plateau $R_S^A(x_p<x<x_{eq},Q_0^2)=R_S^A(x_p,Q_0^2)$.
Now the approximate plateau in $R_S^A$ controls the height of the
anti-shadowing peak in $R_V^A$. At $x=x_{eq}$ the sea-quark ratio 
$R_S^A(x_p,Q_0^2)$ becomes equal to $R_{F_2}^A(x_p,Q_0)$,
and at $x>x_{eq}$, in lack of further information on $R_S^A$, 
we again use the simplest approximation 
$R_V^A= R_S^A= R_{F_2}^A$. The precise values of $x_p$ and $x_{eq}$, 
(i.e. the location and height of the plateau in $R_S^A$), together with 
the parameters  for $R_V^A$ are  first constrained by the baryon number 
conservation at $Q^2=Q_0^2$ and after the DGLAP evolution by the DY data 
\cite{E772}.

Finally, we define the nuclear gluon ratio by
\begin{equation}
R_G^A(x,Q^2) \equiv g_A(x,Q^2)/g(x,Q^2),
\label{RG}
\end{equation}
and specify $R_G^A(x,Q_0^2)$ to have the full input for the
DGLAP evolution. We constrain $R_G^A(x,Q_0^2)$ with
momentum conservation \cite{FSL,KJE}:
\begin{eqnarray}
1 =
\int_0^1 dx\, x \biggl\{ 
g(x,Q_0^2)R_G^A(x,Q_0^2) + [u_V(x,Q_0^2) +d_V(x,Q_0^2)]R_V^A(x,Q_0^2)+ 
\nonumber
\\
2[\bar u(x,Q_0^2) + \bar d(x,Q_0^2) + s(x,Q_0^2)]R_S^A(x,Q^2)
 \biggr\},
\label{MOMENTUM}
\end{eqnarray}
where $R_V^A$ and $R_S^A$ are determined as described above.
Compared to free nucleons, we find that some momentum
is transferred from quarks to gluons in nuclei.
This effect is not very large: in $A\!=\!208$ the glue carries
about 4~\% more momentum than in a free nucleon.
This is in agreement with the earlier studies \cite{FSL,KJE}.
Consequently, without any $x$-dependence in $R_G^A(x,Q_0^2)$
we would have $R_G^A(x,Q_0^2)=1.04$. 

Since the sea quarks are shadowed at small $x$, we expect
shadowing of the nuclear gluons as well.  A requirement of
stable scale evolution can be used together with the recent NMC data
\cite{NMC96} to further constrain nuclear gluon shadowing. At the
small-$x$  limit of the DGLAP equations one obtains
\begin{eqnarray}
\frac{\partial R_{F_2}^A(x,Q^2)}{\partial \log Q^2}
&=&
\frac{ {\partial F_2^D(x,Q^2)} / {\partial \log Q^2} }{F_2^D(x,Q^2)}
\biggl\{\frac{ {\partial F_2^A(x,Q^2)} / {\partial \log Q^2} }
           { {\partial F_2^D(x,Q^2)} / {\partial \log Q^2} }
- R_{F_2}^A(x,Q^2)\biggr\} \\
&\approx&
\frac{5\alpha_s}{9\pi}\frac{xg(2x,Q^2)}{F_2^D(x,Q^2)}
\biggl\{R_G^A(2x,Q^2)-R_{F_2}^A(x,Q^2)\biggr\},
\label{SAT}
\end{eqnarray}
where we have used the result
${\partial F_2(x,Q^2)}/{\partial \log Q^2} \approx {5\alpha_s}
xg(2x,Q^2)/ {9\pi}$, derived in \cite{PRYTZ}.
As for the initial $R_{F_2}^A(x,Q_0^2)$, we will also assume a
saturation in shadowing of the glue, so that
$R_G^A(2x,Q_0^2)\approx R_G^A(x,Q_0^2)$ at the limit $x\rightarrow0$.
For the GRV-LO set we are using, the ratio $xg(2x,Q_0^2)/F_2(x,Q_0^2)$ 
is non-zero and grows slowly at small values of $x$, so $R_{F_2}^A$
evolves towards $R_G^A$ due to the factor $R_G^A-R_{F_2}^A$ in
Eq.~(\ref{SAT}). An initial condition stable in the evolution
is obtained by requiring $R_G^A(x,Q_0^2)\approx R_{F_2}^A(x,Q_0^2)$ 
at very small values of $x$.

Due to momentum conservation, the loss of gluons in the shadowing 
region must lead to an increase of gluon distribution somewhere
at larger $x$. In Ref.~\cite{KJE}, it was assumed that $R_G^A(x,Q^2)$ 
is constant at large $x$. This, however, resulted in somewhat unstable
profiles for $R_G^A$ and $R_S^A$ in the evolution. Therefore, we expect 
that there is an EMC-effect also for the glue, and that $R_G^A(x,Q_0^2)$
is peaked around the same values as  $R_V^A(x,Q_0^2)$ and $R_S^A(x,Q_0^2)$.

In practice, we determine the gluon ratio as follows.  We start with
$R_G^A(x,Q_0^2)\approx R_{F_2}^A(x,Q_0^2)$ at small values of $x$.  
Should we use this equality for all values of $x$, some momentum would 
be missing for all nuclei (11~\% for $A\!=\!208$).  This indicates 
that conservation of momentum requires quite strong anti-shadowing  
for the gluons.  For $R_G^A(x,Q_0^2)$ we use the functional form given  
in the Appendix. In determination of the position and width of the 
anti-shadowing peak (which we assume to be independent of $A$) we use 
Ref.~\cite{PIRNER} to constrain the value of $x$ where 
$xg_{\rm Sn}/xg_{\rm C}\approx 1$. The amount of anti-shadowing then 
follows from momentum conservation, leading to a good agreement with 
Ref.~\cite{PIRNER} once the position and width parameters of the 
peak in $R_G^A$ are fixed.

\begin{figure}[tb]
\vspace{-3cm}
\centerline{\hspace*{0cm} \epsfxsize=18cm\epsfbox{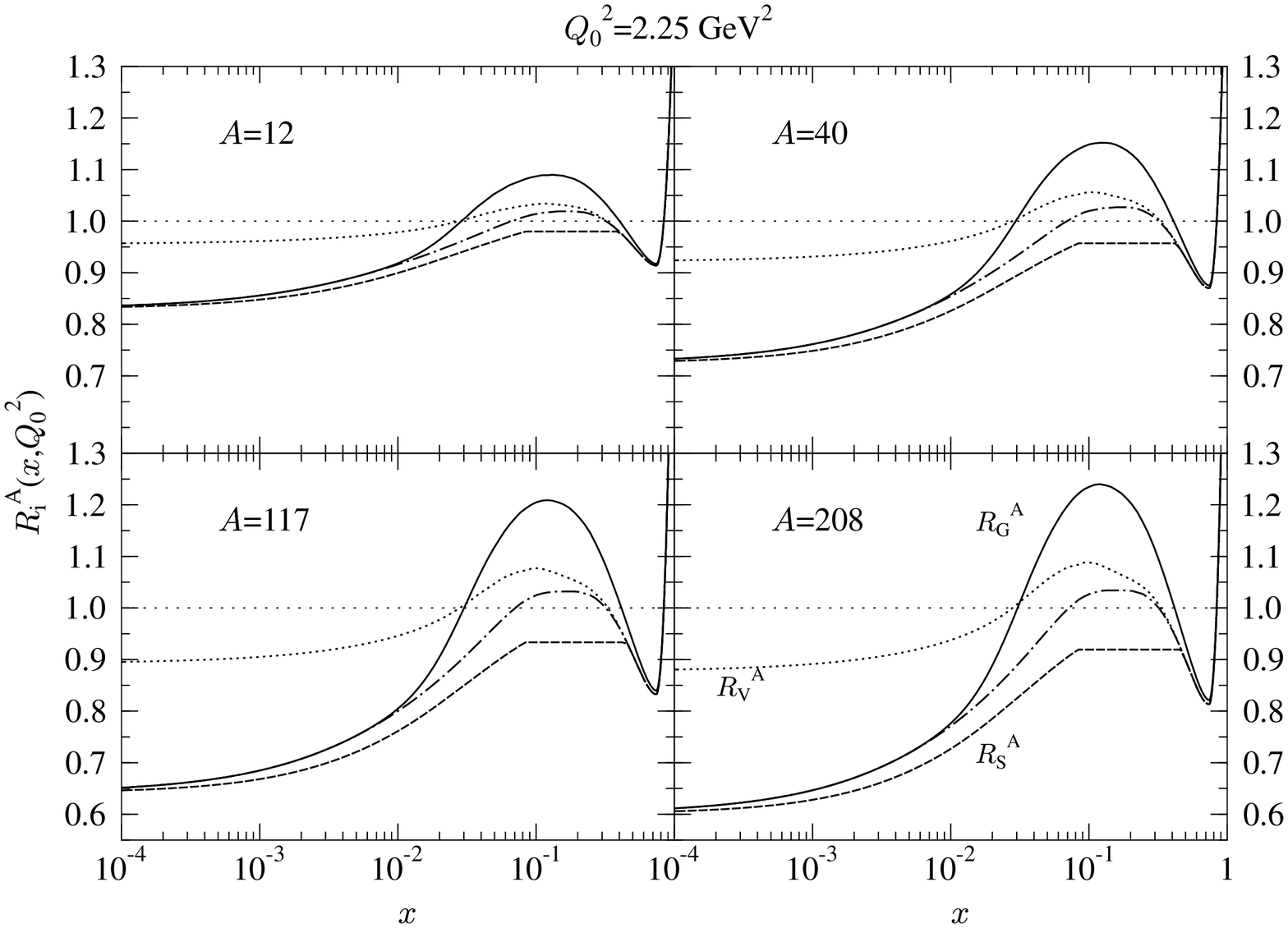}}
\vspace{-1cm}
\caption[a]
{{
\small The initial nuclear ratios $R_G^A(x,Q_0^2)$ (solid line), 
$R_V^A(x,Q_0^2)$ (dotted) and $R_S^A(x,Q_0^2)$ (dashed) for isoscalar 
nuclei at $Q_0^2=2.25$~GeV$^2$. The ratio $R_{F_2}^A(x,Q_0^2)$ 
(dotted-dashed) is also shown.
}}
\label{RQ0}
\end{figure}

The initial nuclear ratios $R_G^A(x,Q_0^2), R_V^A(x,Q_0^2)$ and 
$R_S^A(x,Q_0^2)$ at $Q_0^2 = 2.25$~GeV$^2$ are shown for isoscalar 
nuclei in Fig.~\ref{RQ0} together with the ratios $R_{F_2}^A(x,Q_0^2)$. 
Correlations in these ratios are easily understood from the figures: 
the more $R_V^A$ is shadowed, the more it has anti-shadowing, 
and the more $R_S^A$ is suppressed in the anti-shadowing region.
Similarly, due to momentum conservation, the more shadowing $R_G^A$ has, the
more anti-shadowing is needed. As seen in the figures, there is
more anti-shadowing and less shadowing for $R_V^A$ than for $R_{F_2}^A$.
Correspondingly, $R_S^A$ is more shadowed and not anti-shadowed at all at
$Q_0^2 = 2.25$~GeV$^2$. The gluon ratios $R_G^A$ turn out to be clearly
more anti-shadowed than $R_{F_2}^A$ and $R_V^A$.
In comparison with the earlier studies, the lack of sea-quark
anti-shadowing agrees with \cite{FSL,KJE}. We get, however, somewhat 
more anti-shadowing for gluons than in \cite{FSL,KJE}, which is a 
consequence of the small-$x$ enhancement of gluon densities in 
a proton.

\section{Scale evolution and results}

Scale evolution is straightforward to carry out
once the parton fusion corrections \cite{GLRMQ} are neglected.
In the HERA data for $F_2^p(x,Q^2)$ \cite{HERA,HERAlowQ} there is no
evidence of the fusion corrections at $Q^2\gsim 1$~GeV$^2$ and
$x\gsim 10^{-4}$. For nuclei, parton fusion should be stronger due to 
its expected $A^{1/3}$ scaling \cite{GLRMQ,EQW}, but since we start 
the evolution at $Q_0^2=2.25$~GeV$^2~>~1$~GeV$^2$, the effects of 
these corrections should be small \cite{KJE}, at least in the $x$ 
range of the NMC data \cite{NMC96}.
The role of the fusion corrections is, however, a very
interesting question \cite{QIU,KJE,KUMANO,EQW} which deserves
further analyses with constraints for the proton
from the new HERA data \cite{HERAlatest}
at very small $Q^2$ and very small $x$.

\begin{figure}[tb]
\vspace{-.5cm}
\centerline{\epsfxsize=14.cm \epsfbox{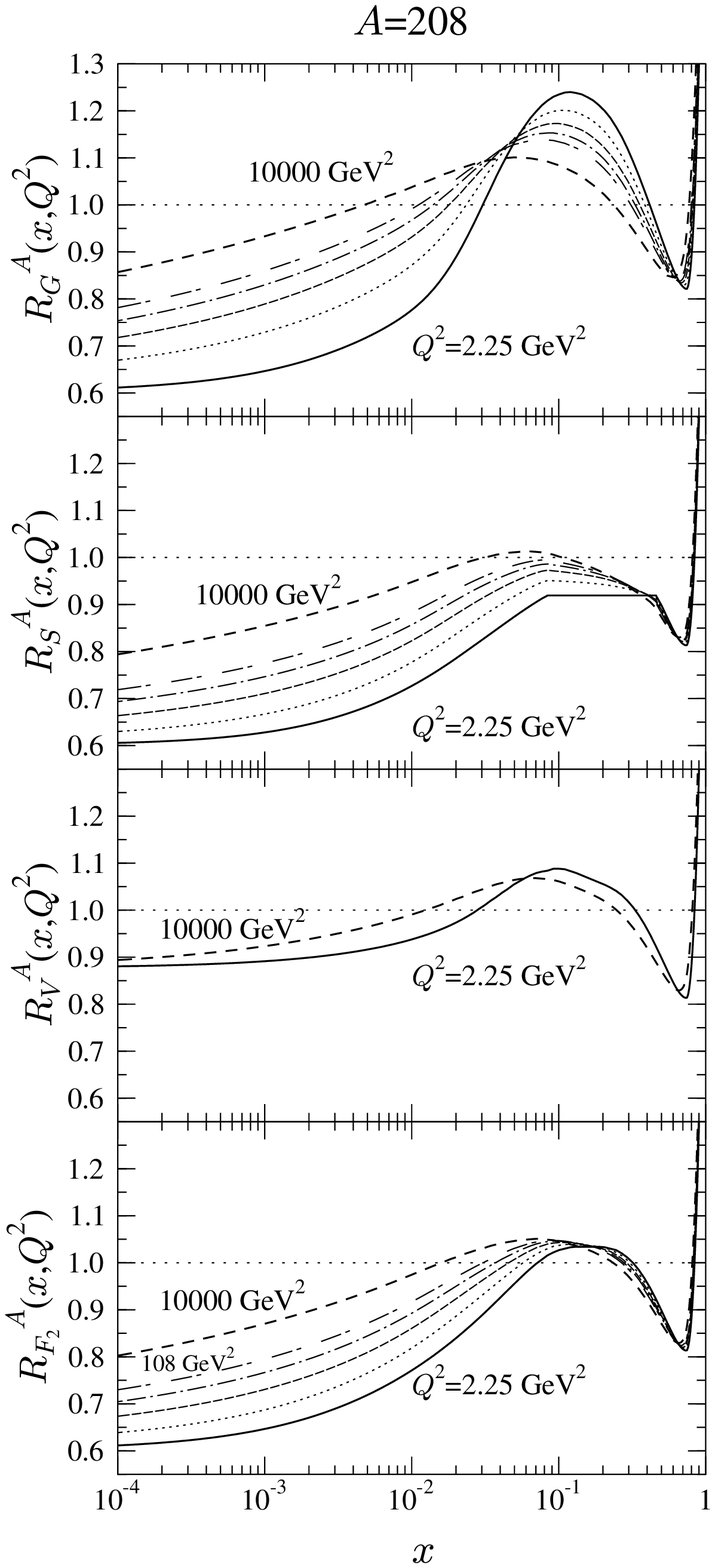}}
\vspace{-0.5cm}
\caption[a]
{{\small 
Scale evolution of the ratios $R_G^A(x,Q^2)$, $R_S^A(x,Q^2)$, 
$R_V^A(x,Q^2)$ and  $R_{F_2}^A(x,Q^2)$ for an isoscalar nucleus
$A$=208. The ratios are shown as functions of $x$ at fixed 
values of $Q^2=$ 
2.25~GeV$^2$ (solid lines),
5.39~~GeV$^2$ (dotted),
14.7~~GeV$^2$ (dashed),
39.9~~GeV$^2$ (dotted-dashed),
108~GeV$^2$ (double-dashed), equidistant in $\log Q^2$, and 
10000~GeV$^2$ (dashed). For $R_V^A$ only the first and 
last ones are shown.
}}
\label{RQ2}
\end{figure}

We use the parametrization of GRV-LO \cite{GRVLO} for the initial
parton distributions of free protons. We evolve these and the
initial nuclear distributions from $Q_0^2 = 2.25$~GeV$^2$
to $Q^2 \sim 10000$~GeV$^2$ by using the DGLAP equations for gluons and
valence and sea quarks of each flavor. Massive quarks are generated
through the evolution above each mass threshold as in \cite{GRVLO}.
We solve the evolution equations directly in ($x,Q^2$)-space.
The resulting scale evolution of the ratios $R_G^A$, $R_S^A$ and
$R_V^A$ is shown in Figs.~\ref{RQ2} for $A\!=\!208$. Compared 
with an earlier study \cite{KJE}, the difference in the evolution of 
$R_G^A$ and $R_{F_2}^A$ is not as dramatic, due to the enhanced,
more stable gluon and sea-quark distributions at small vales 
of $x$.

\begin{figure}[t]
\vspace{-1.5cm}
\centerline{\hspace*{0cm} \epsfxsize=14cm\epsfbox{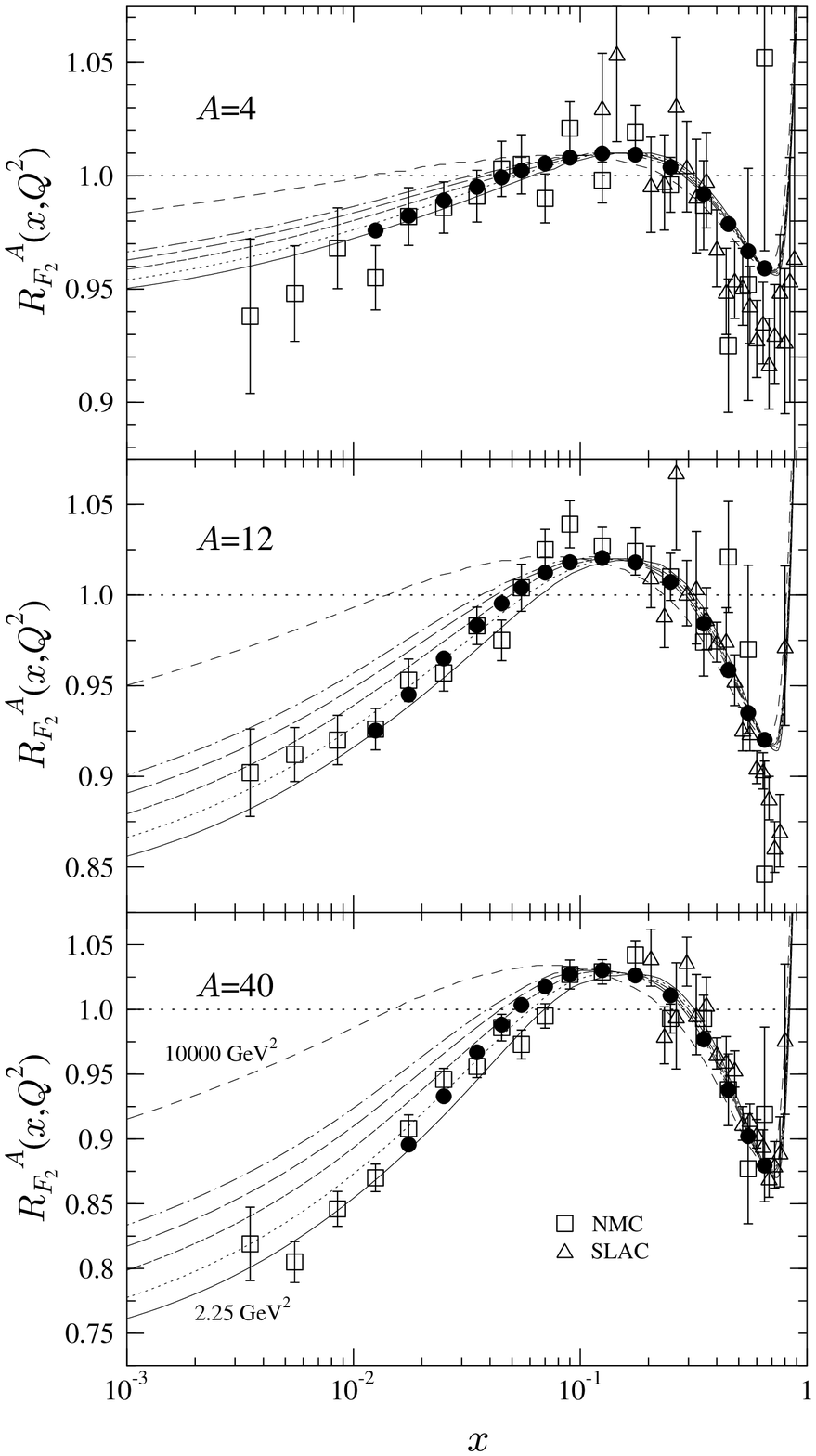}}
\vspace{-0.5cm}
\caption[a]
{ {\small 
Scale evolution of the ratio $R_{F_2}^A(x,Q^2)$ for isoscalar nuclei 
$A$=4, 12 and 40. As in Fig.~\ref{RQ2}, the ratios are plotted as 
functions of $x$ but with $Q^2$ fixed to 
2.25, 3.70, 6.93, 12.9, 24.2~GeV$^2$, equidistant in $\log Q^2$, 
 and 10000~GeV$^2$. The reanalyzed NMC data \cite{NMCre} is shown 
by the boxes, the reanalyzed SLAC data by the triangles \cite{SLACre}.
The statistical and systematic errors have been added in quadrature.
The filled circles show our calculation at the $\langle Q^2 \rangle$ values 
of the NMC data. Notice that the vertical scale of each panel is different.
}}
\label{RF2Q2}
\end{figure}

In Figs.~\ref{RF2Q2} we plot the scale evolution of the ratio
$R_{F_2}^A(x,Q^2)$ for isoscalars $^4_2$He, $^{12}_{~6}$C, $^{40}_{20}$Ca
and for an artificial isoscalar nucleus $A\!=\!208$ at different fixed $Q^2$.
The data shown and used in the analysis, are the re-analyzed data from 
NMC \cite{NMCre} and SLAC \cite{SLACre}. For a more transparent 
comparison, we show our calculation with filled circles at $Q^2$ equal to
the $\langle Q^2\rangle$ of the NMC data at different $x$. 

\begin{figure}[ht]
\vspace{-3cm}
\centerline{\hspace*{0cm} \epsfxsize=14cm\epsfbox{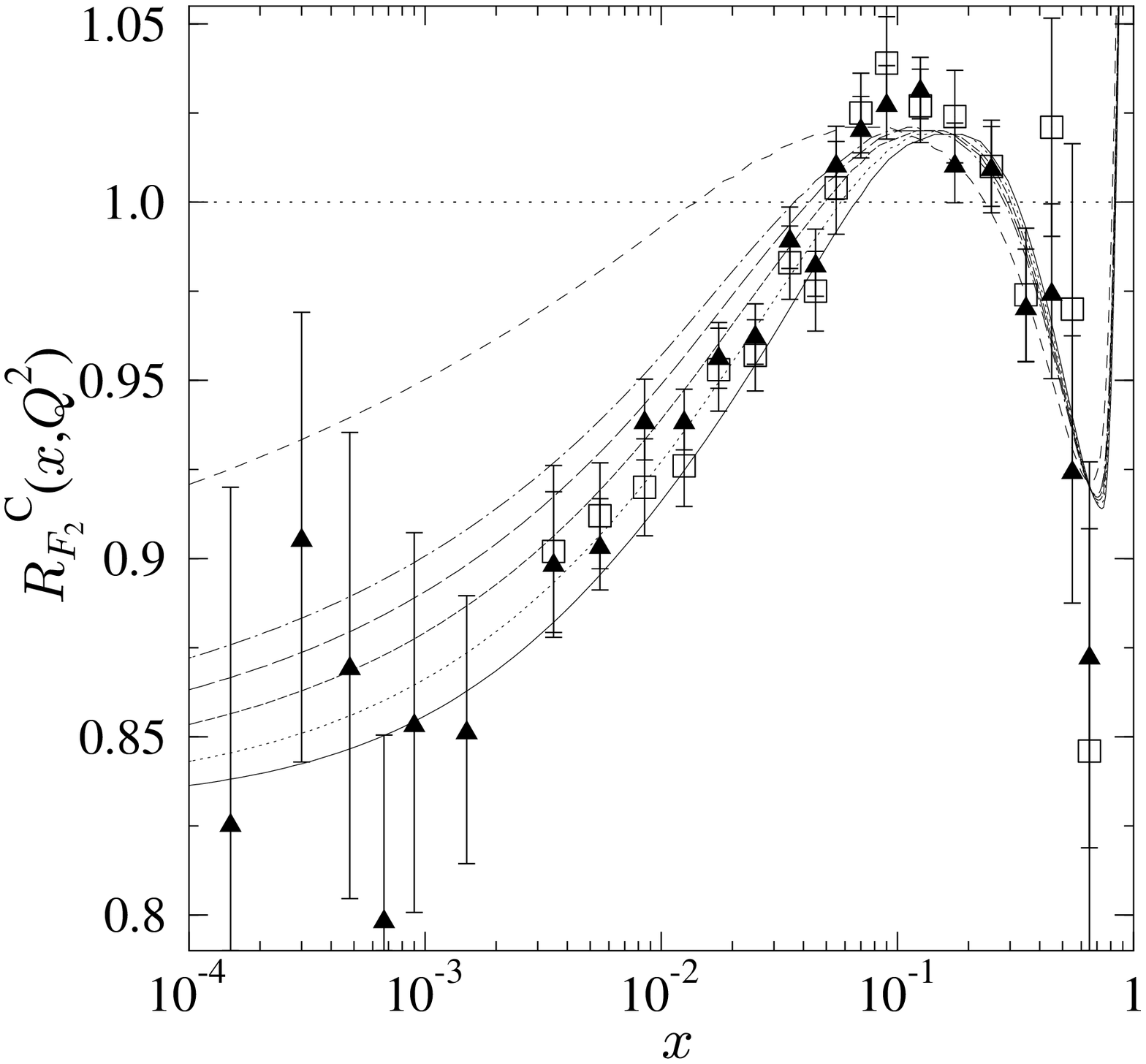}}
\vspace{-1cm}
\caption[a]
{ {\small 
The ratio $R_{F_2}^{\rm C}(x,Q^2)$ shown together with the 
renanalyzed NMC data \cite{NMCre} (boxes) and the combined
NMC data \cite{NMCsat} (triangles). The calculated  
$R_{F_2}^{\rm C}$ are shown at the same fixed values of 
$Q^2$ as in Fig.~\ref{RF2Q2}. Notice that at $x<0.01$ the 
scales $\langle Q^2\rangle$ of the data are less than our 
$Q_0^2=2.25$~GeV$^2$. The statistical and systematic 
errors of the data are added in quadrature.
}}
\label{RF2x}
\end{figure}

In Fig.~\ref{RF2x}, we show the behaviour of $R_{F_2}^A(x,Q^2)$ at very
small $x$, together with the NMC data \cite{NMCsat} and \cite{NMCre} for 
$^{12}$C. This figure confirms the consistency of our initial assumption 
of the saturation of shadowing in $R_{F_2}^A(x,Q_0^2)$: at fixed $x$, 
shadowing decreases with increasing $Q^2$ and therefore our initial 
$R_{F_2}^A(x,Q_0^2)$ should not lie below the data measured at $Q^2<Q_0^2$.
This of course implies an implicit assumption that the scale dependence 
of $R_{F_2}^A$ in the nonperturbative region is towards the same 
direction as in the perturbative region. It should be kept in mind, 
however, that at $Q^2>Q_0^2$ the sign of 
$\partial R_{F_2}^A(x,Q^2)/\partial Q^2$, as indicated by Eq.~(\ref{SAT}), 
actually depends on the gluon shadowing at small $x$, i.e. if the gluons 
were more shadowed, the decrease of shadowing in  $R_{F_2}^A$ would be 
slower, or shadowing could even increase. In this case $R_{F_2}^A(x,Q_0^2)$ 
should  be re-determined accordingly. In this way, as pointed out in 
\cite{KJE}, at small values of $x$ the scale evolution of $R_{F_2}^A$ 
reflects the amount of nuclear gluon shadowing.

\begin{figure}[ht]
\vspace{-3cm}
\centerline{\hspace*{0cm} \epsfxsize=14cm\epsfbox{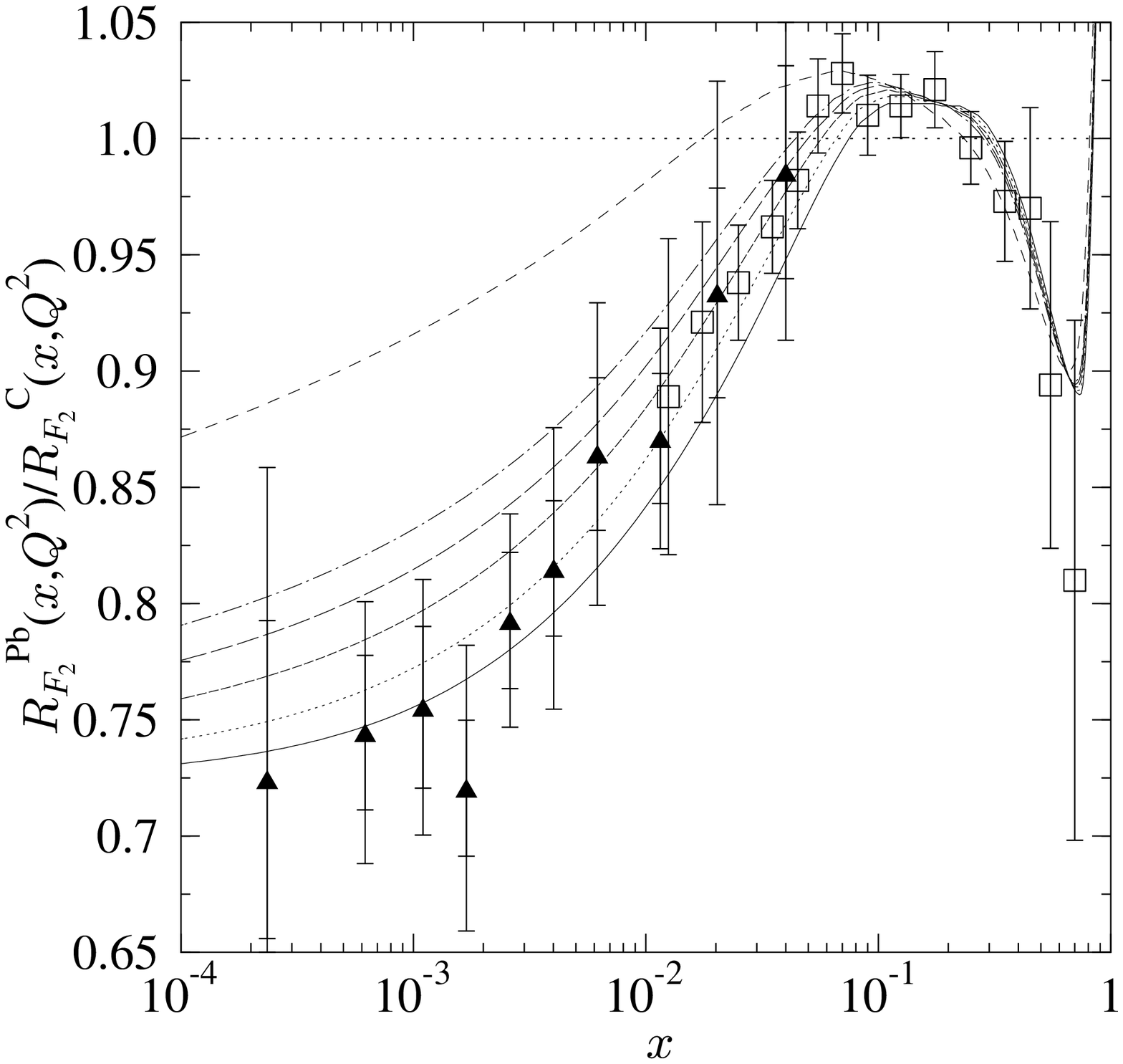}}
\vspace{-1cm}
\caption[a]
{ {\small 
The ratio $R_{F_2}^{\rm Pb}(x,Q^2)/R_{F_2}^{\rm C}(x,Q^2)$ as a function 
of $x$ at the same fixed values of $Q^2$ as in  Fig.~\ref{RF2Q2}.
The NMC data for Pb/C \cite{NMCsys} is shown by the squares and 
with statistical and systematic errors added in quadrature.
The ratio of the E665 data for $R_{F_2}^{\rm Pb}$ and $R_{F_2}^{\rm C}$
\cite{E665} is shown with the triangles. The inner error bars
are obtained by including the independent statistical errors only,
and the outer ones by adding first the statistical and systematic 
errors separately for Pb/D and C/D in quadrature, and then taking these
errors to be independent. This figure shows how our calculation relates
to the small-$x$ region where the measurements are at nonperturbative 
scales. 
}}
\label{F2E665}
\end{figure}

Small-$x$ data for $R_{F_2}^A$ exist also from the E665 collaboration 
\cite{E665}, but they lie above the NMC data (see e.g. \cite{NMCsys}).
Because of the smaller error bars in the NMC data, we decided not
to use the absolute E665 data for determining $R_{F_2}^A(x,Q_0^2)$ here.
However, as pointed out
in \cite{NMCsys}, if one considers the {\em ratios} of different
nuclei, the two data sets are consistent. Therefore, we have
also used the ratio $F_2^{\rm Pb}/F_2^{\rm C}$ as obtained
from the E665 data \cite{E665}. In Fig.~\ref{F2E665} we show the 
results at small $x$ as in Fig.~\ref{RF2x} but for the ratio of 
ratios, $R_{F_2}^{\rm Pb}/R_{F_2}^{\rm C}$ with the data from 
\cite{NMCsys} and \cite{E665}. Note that the inner (outer) error 
bars under(over)estimate the real errors of the E665 data (see 
the figure caption). Even without a more precise 
$\chi^2$ minimization, this figure partly helps us in constraining 
the $A$ dependence at small values of $x$ in the parametrization of 
$R_{F_2}^A$ given in the Appendix.

\begin{figure}[tb]
\vspace{-1.5cm}
\centerline{\epsfxsize=16cm\epsfbox{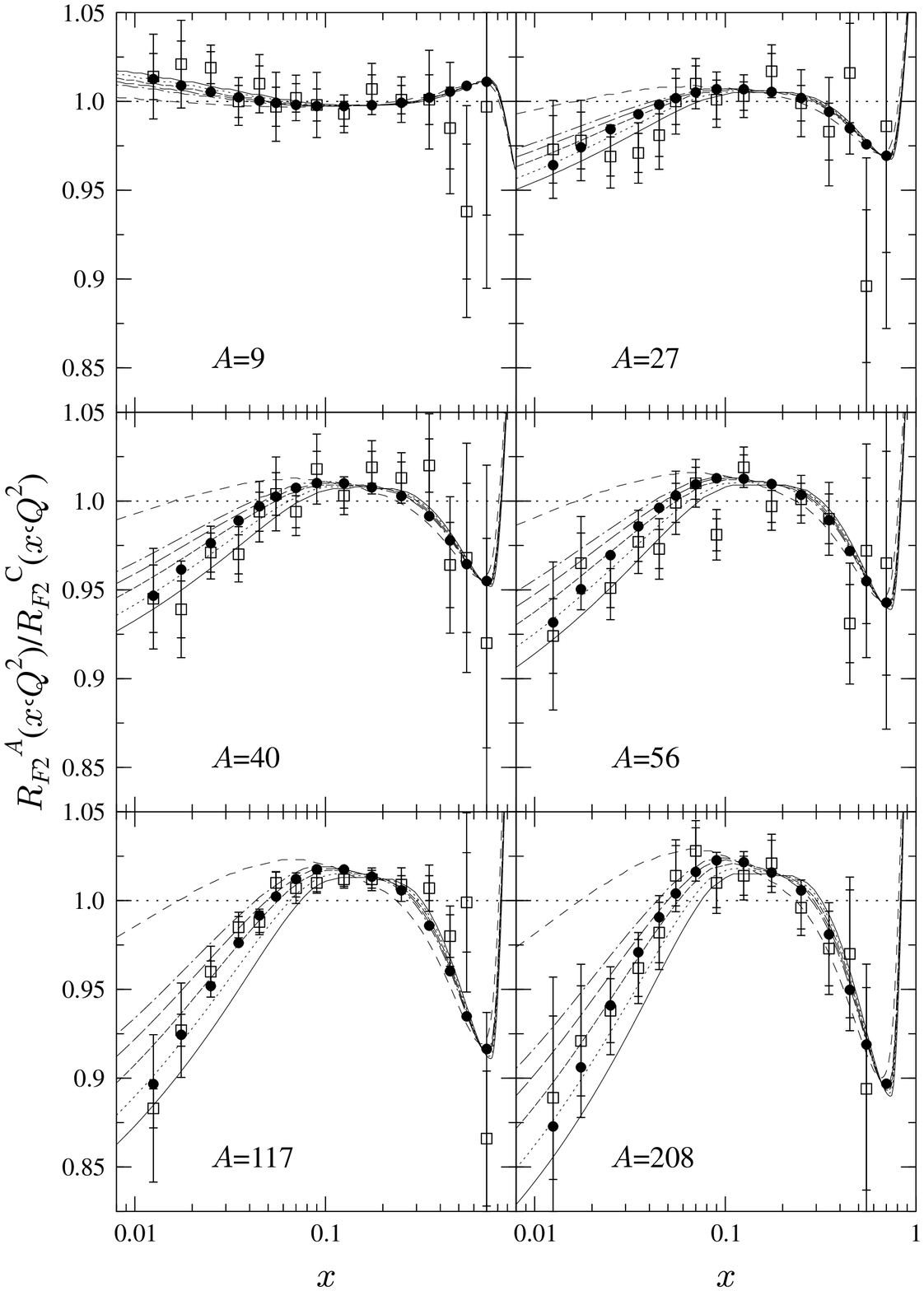}}
\vspace{-.5cm}
\caption[a]
{ {\small 
The ratios $R_{F_2}^A(x,Q^2)/R_{F_2}^C(x,Q^2)$ as functions of $x$ for 
isoscalar nuclei. The values of $Q^2$ are the same as in Fig.~\ref{RF2Q2}.
For comparison with the NMC data \cite{NMCsys} (boxes), the filled 
circles show our calculation at the $\langle Q^2 \rangle$ of the data. 
The inner error bars stand for the statistical errors only, the outer 
ones for statistical and systematic errors added in quadrature.
}}
\label{A1A2}
\end{figure}

The systematics in nuclear mass number $A$, especially at $x\lsim 0.1$,
is presented by the NMC collaboration in \cite{NMCsys}, and we have 
also made use of these data. In Figs.~\ref{A1A2} we show the scale
evolution of $F_2^A/F_2^{\rm C}$, i.e. of $R_{F_2}^A/R_{F_2}^{\rm C}$,
and the comparison with the NMC data. Since the data are  
(approximatively) corrected for non-isoscalar effects, we have always 
$A\!=\!2Z$. In our analysis, the data for the $(A\!=\!117)$/C ratio 
gives the most stringent constraints. Again, for an easier comparison, 
the filled symbols show our calculation at the $\langle Q^2\rangle$ 
of the data at different $x$.

\begin{figure}[tb]
\vspace{-0.5cm}
\centerline{\hspace*{0cm} \epsfxsize=17cm\epsfbox{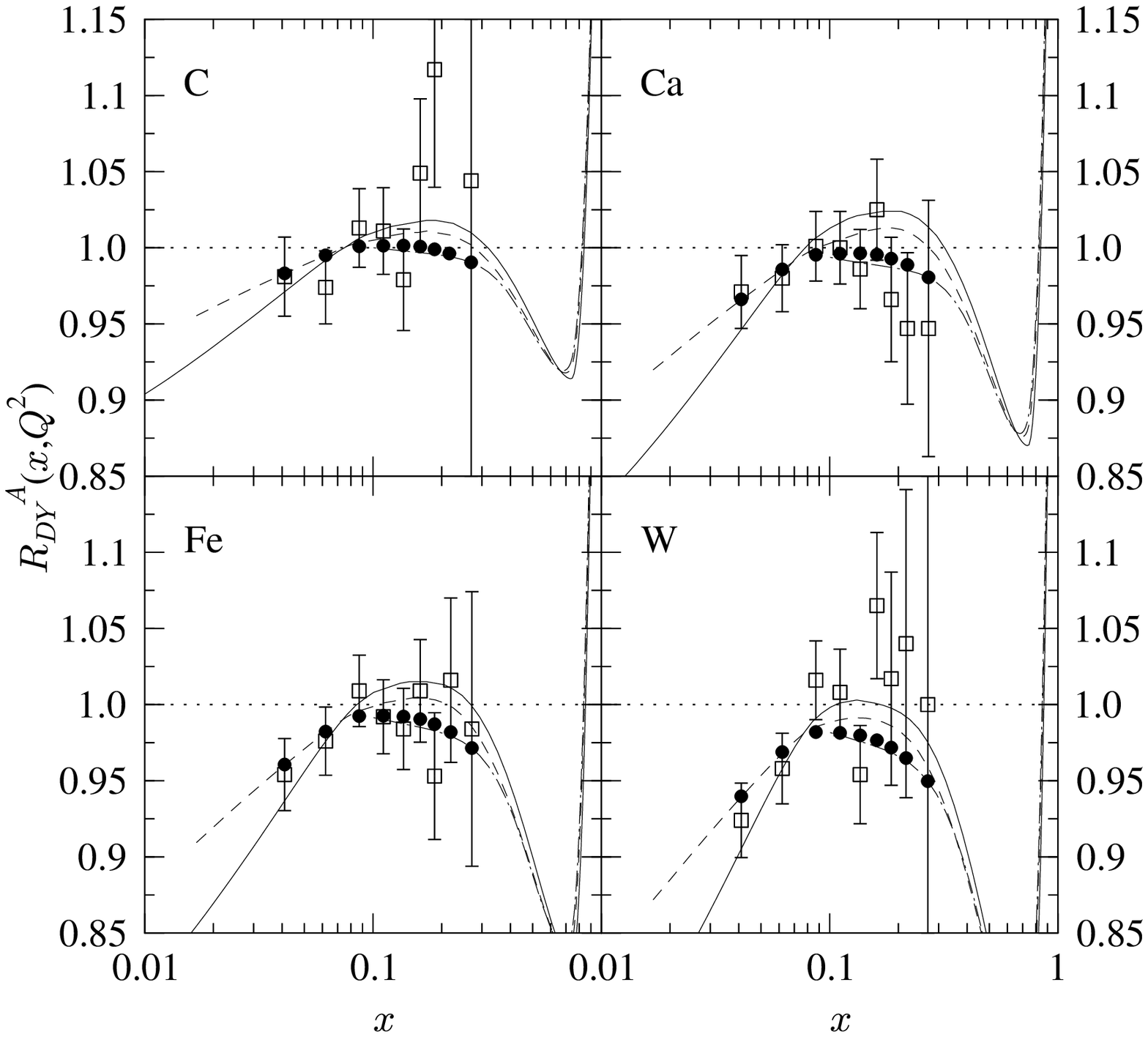}}
\vspace{-0cm}
\caption[a]
{{\small 
The ratios of differential Drell-Yan cross sections in 
$pA$ and $pD$ as functions of $x=x_2$ for $^{12}_{~6}$C$/D$, 
$^{40}_{20}$Ca$/D$, $^{56}_{26}$Fe$/D$ and $^{184}_{~74}$W$/D$. 
Our calculation for $R_{DY}^A(x,Q^2)$ of Eq.~(\ref{RDY}) is 
shown at fixed values of the invariant mass $Q^2=$ 2.25~GeV$^2$ 
(solid line), 24.2~GeV$^2$ (dashed), and 139~GeV$^2$ (dotted-dashed).
The data shown by the boxes is from E772 \cite{E772}. In the graph the 
statistical errors and the quoted 2~\% systematic errors are added 
in quadrature. The filled circles show our calculation 
at the $\langle Q^2\rangle$ of the data \cite{PMcG}.
}}
\label{DY}
\end{figure}

How the Drell--Yan data from the E772-collaboration \cite{E772} enters
our analysis, can be seen from Figs.~\ref{DY} where we have plotted
the ratio of Drell--Yan cross sections as defined in Eq.\ (\ref{RDY}) at
different invariant masses $Q^2$ for $^{12}_{~6}$C, $^{40}_{20}$Ca,
$^{56}_{26}$Fe and $^{184}_{~74}$W. The non-isoscalar terms for Fe and W 
have been taken into account according to Eq.~(\ref{RDY}), 
by using the corrections obtained with isoscalar nuclei. 
As before, the filled symbols show our calculation at the mass 
values $\langle Q^2\rangle$ of the data \cite{PMcG}.
The constraint for determining the initial ratios $R_S^A(x,Q_0^2)$
and $R_V^A(x,Q_0^2)$ enters as follows: let us suppose that
valence quarks are shadowed less than in Figs.~\ref{RQ0}.
This leads to a smaller anti-shadowing for $R_V^A(x,Q_0^2)$ 
because of baryon number conservation,  Eq.~(\ref{CHARGE}). 
At the same time, Eq.\ (\ref{RF2}) brings $R_S^A(x,Q_0^2)$
closer to $R_{F_2}^A$ at $x\sim 0.1$,  i.e. $R_S^A$ becomes less
suppressed. In the scale evolution, there is a cross-over region in
$R_{DY}^A$ where evolution is very weak (see Figs.~\ref{DY}),
and the location of which depends on the plateau in $R_S^A(x\sim 0.1,Q_0^2)$.
With less suppressed $R_S^A$ at $x\sim 0.1$, the location of the
cross-over  region shifts to larger $x$. Consequently, the evolution of
$R_{DY}^A$ becomes too fast at small values of $x$, and the
calculated values overshoot the data. On the other hand, if $R_V^A$
is much more shadowed than in Figs.~\ref{RQ0},
the cross-over region in $R_{DY}^A$ shifts to the left,
scale evolution becomes too fast at $x\gsim 0.1$, and the calculation
falls below the data. Iterating between these possibilities we have
obtained the initial $R_S^A(x,Q_0^2)$ shown in the figures.

\begin{figure}[ht]
\vspace{-3cm}
\centerline{\hspace*{0cm} \epsfxsize=14cm\epsfbox{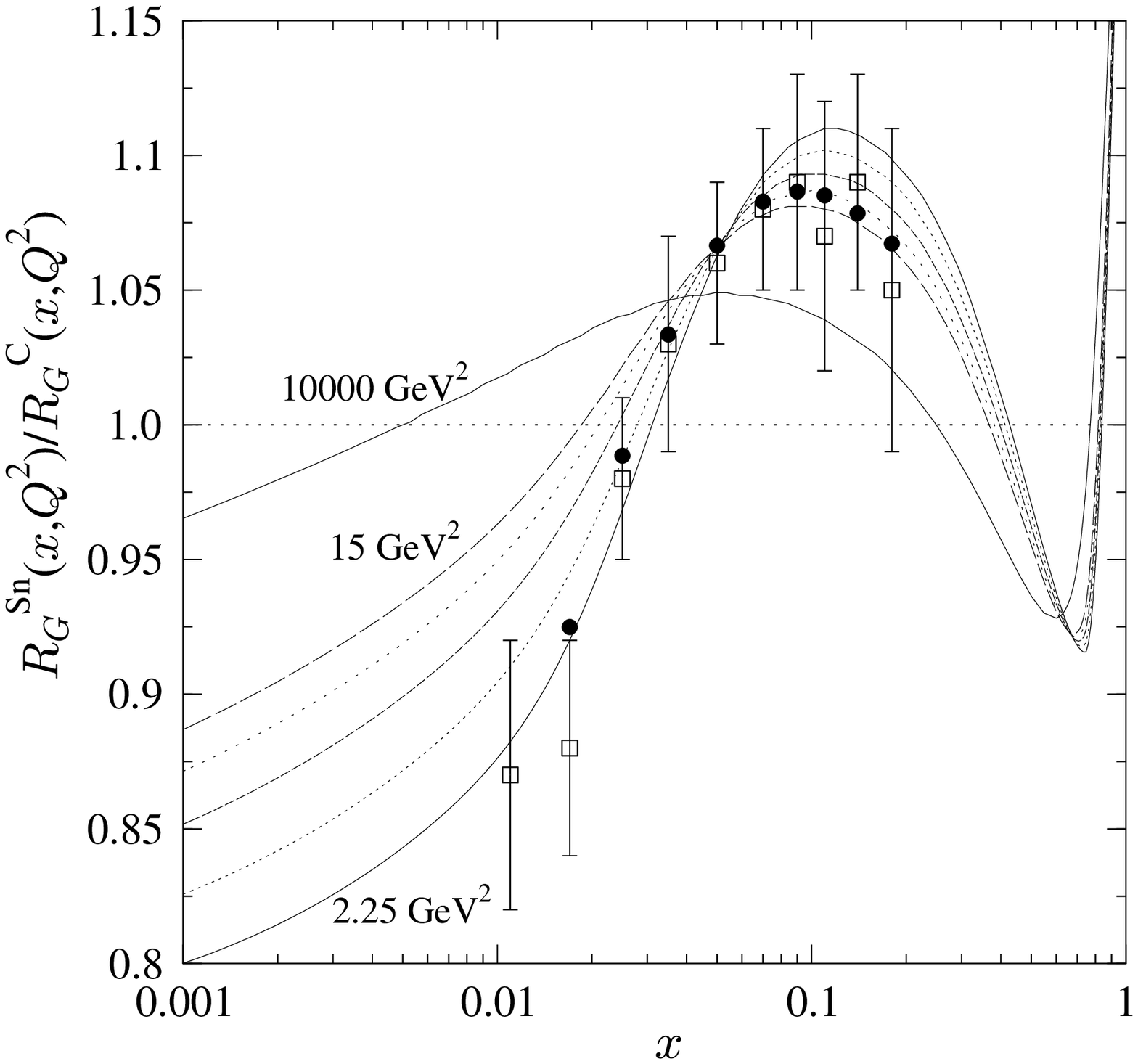}}
\vspace{-0cm}
\caption[a]
{ {\small 
The gluon ratios $R_G^{\rm Sn}(x,Q^2)/R_G^{\rm C}(x,Q^2)$ as functions of $x$
at fixed values of $Q^2\!=$~2.25, 3.27, 5.39, 8.89, 14.7~GeV$^2$, equidistant in 
$\log Q^2$,  and 10000~GeV$^2$. The filled circles show the comparison
with the results of Ref.~\cite{PIRNER}, presented by the boxes.
}}
\label{Rpirner}
\end{figure}

In Fig.~\ref{Rpirner}, comparison with the results of
Ref.~\cite{PIRNER} for the gluon ratio $g_{\rm Sn}/g_{\rm C}$ is shown.
Again, the curves are plotted at fixed values of $Q^2$ and the filled 
symbols show our calculation at the $\langle Q^2\rangle$ of the preliminary 
NMC data \cite{MUCKLICH} as quoted in \cite{PIRNER}.  It should be noted
that even though only the point where $g_{\rm Sn}/g_{\rm C}\sim 1$
was constrained to agree with \cite{PIRNER}, the resulting amount
of shadowing and anti-shadowing comes out in a very good agreement 
with the results of Gousset and Pirner.

\begin{figure}[tb]
\vspace{-1cm}
\centerline{\hspace*{0cm} \epsfxsize=16cm\epsfbox{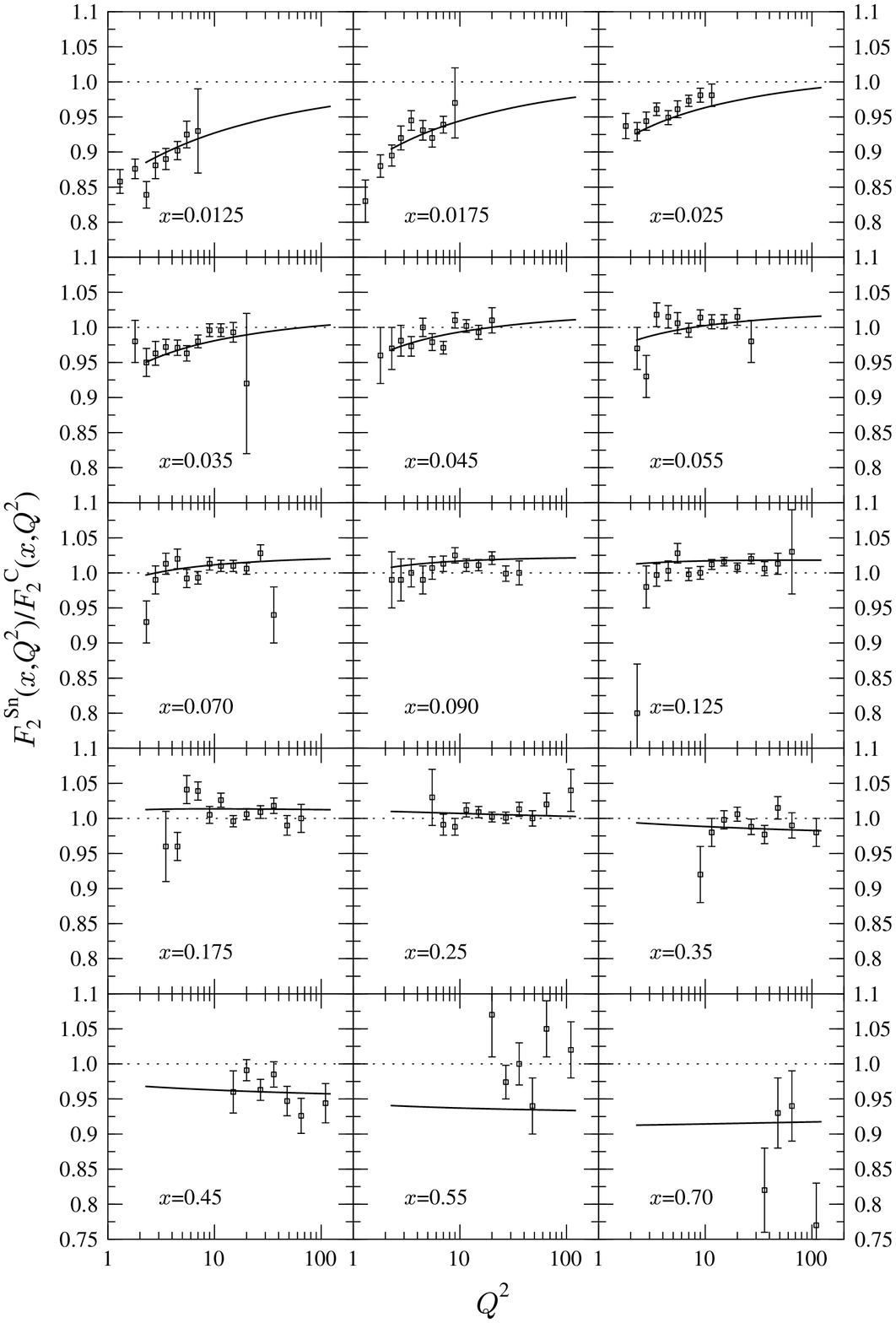}}
\vspace{-0.cm}
\caption[a]
{ {\small 
The calculated scale evolution of $F_2^{\rm Sn}(x,Q^2)/F_2^{\rm C}(x,Q^2)$  
compared with the NMC data \cite{NMC96} at different fixed values of $x$.
The data are plotted with statistical errors only.
}}
\label{SnC}
\end{figure}

Finally, in Figs.~\ref{SnC}, we present the main result of our study,
comparison of the calculated scale  evolution of $F_2^{\rm Sn}/F_2^{\rm C}$
with the recent NMC data \cite{NMC96}. The data is plotted with statistical
errors only because we are more interested in the slopes of $Q^2$ evolution 
than in the absolute normalization to the data. Notice that normalization 
of our curves at each fixed $x$ in Figs.~\ref{SnC} depends on the initial 
ratio shown for Sn/C in Fig.~\ref{A1A2}. In fact we obtain a fairly good 
agreement also for the overall normalization of the data. The changes in 
the sign of $\partial (F_2^{\rm Sn}/F_2^{\rm C})/\partial Q^2$ do not seem 
to be in contradiction with the data, either. It is also interesting to 
notice that at small values of $x$ the slope is not linear in $\log Q^2$.

After performing the scale evolution, let us look back into the 
assumptions made in constraining the nuclear ratios at the initial 
scale $Q_0^2$.
As expected on the basis of Eq.\ (\ref{SAT}), the approximation 
$R_G^A = R_{F_2}^A$ for the initial distributions at extremely 
small $x$ turns out to be quite stable in the evolution; at the 
smallest $x$ we consider, $x=10^{-6}$, deviations from 
$R_G^A=R_{F_2}^A$ are only about 5~\% up to 
$Q^2 = 10000$~GeV$^2$ for $A\!=\!208$.
For the valence quarks, the initial approximations
$R_{u_V}^A= R_V^A(x,Q_0^2)$ and $R_{d_V}^A= R_V^A(x,Q_0^2)$ are 
very stable: deviations are within 1~\% at $x<0.7$ at all scales 
considered.
For the GRV-LO set \cite{GRVLO}, there is no difference between 
$\bar u$ and $\bar d$, so once the initial approximation 
$R_{\bar u}^A = R_{\bar d}^A$ is made, it holds exactly at all 
scales. The initial approximation $R^A_{\bar u+\bar d}= R_S^A$ 
is also a stable one, the deviations are within 2~\% during the evolution.
Initial approximation $R_{\bar s}^A = R_S^A$ is slightly less 
stable, but deviations do stay within 7~\% at all scales considered.
The deviations are concentrated in the region $x=0.01...0.1$, 
where the nuclear effects for sea quarks are small in any case. 
During the evolution, the strange quarks tend to develop more anti-shadowing 
than $R_S^A$, while the $\bar u$ and $\bar d$ have an opposite tendency 
relative to  $R_S^A$. More work is needed to relax the assumption
$R_s^A(x,Q_0^2)= R_{\bar u+ \bar d}^A(x,Q_0^2)$  but
this additional correction is beyond the scope of this study.

\section{Discussion and Conclusions}

The main emphasis of this study is in the $Q^2$-evolution of 
nuclear effects in parton distributions in the region of small $x$. 
We have used the deep inelastic scattering data from $lA$ collisions
\cite{NMCre}-\cite{E665} and the Drell--Yan dilepton data from $pA$ 
collisions \cite{E772} together with conservation of baryon 
number and momentum to constrain the initial nuclear distributions
as model-independently as possible. As the main result, we have shown 
that a consistent picture arises, and a very good --- almost surprisingly 
good --- agreement with the measured $Q^2$ evolution of the structure 
function ratio $F_2^{\rm Sn}/F_2^{\rm C}$ \cite{NMC96} can be obtained 
already with the lowest order leading twist DGLAP evolution. We can 
therefore conclude that the effect of parton fusion corrections
in the evolution \cite{GLRMQ} is negligible, at least at $x\gsim 0.01$ 
and $Q^2>2.25$~GeV$^2$. In this region, the nuclear modifications are 
effectively built in the nonperturbative initial conditions for the 
DGLAP evolution. This result also agrees with Ref.~\cite{KUMANO}.

We point out, however, that even though we obtain a very good 
agreement with the NMC data \cite{NMC96} and with the analysis 
of Ref.~\cite{PIRNER}, we can confirm our initial assumption of 
gluon shadowing at small values of $x$ only on fairly qualitative 
grounds (stability  of the evolution), rather than through a 
direct comparison with the data. The reason for this is seen 
qualitatively from Eq.~(\ref{SAT}): 
the data on $F_2^{\rm Sn}/F_2^{\rm C}$ at $x\!=\!0.0125$ constrains 
$R_G^A$ at $x\!=\!0.025$, which is only the beginning of the 
gluon shadowing region and where nuclear effects in $R_G^A$ are 
not very strong (see Figs.~\ref{RQ0}). To constrain the gluon 
shadowing further, it would be very important to have data 
for the $Q^2$ dependence of $F_2^{\rm Sn}/F_2^{\rm C}$ available 
at smaller values of $x$. 

The NMC data on deep inelastic $J/\Psi$-production in Sn/C 
\cite{JPSI} is not included in our analysis but it is interesting 
to notice, as pointed out in \cite{PIRNER}, that the amount of 
gluon anti-shadowing in the ratio Sn/C agrees with the excess 
reported in \cite{JPSI}. The gluon anti-shadowing we obtain is  
also consistent with the E789 data on $D$ meson production 
in $pA$ collisions \cite{E789D}, although the error bar on the data 
point is quite large. With the strongly interacting final states,
higher twist production mechanisms \cite{HOYER} may well make the
picture more complicated regarding factorization. However, if the 
initial state effects can be factorized into the nuclear parton 
densities \cite{SALGADO}, our analysis on $R_G^A$ should be of 
direct use also for computing strongly interacting final states in 
nuclear collisions. Our results should also provide more insight 
in understanding the recently measured increase in $J/\Psi$ suppression 
relative to the Drell--Yan background in central Pb-Pb collisions 
at the CERN SPS \cite{NA50}. 

Nuclear structure functions $F_2^{\nu Fe}$, $F_3^{\nu Fe}$ \cite{CCFR} 
are used in the global analyses of parton distributions for 
a {\em free} proton \cite{CTEQ}. Our results should offer a more 
consistent way of unfolding the nuclear effects there.\footnote{We 
thank C.A. Salgado for discussions on this point.} Also the 
coordinate space description of nuclear parton densities 
\cite{VANTTINEN} could be studied in more detail by using the
scale evolved nuclear ratios we have presented here. Related to 
the coordinate space description and scale evolution of nuclear 
parton distributions, the connection of our results with those in
the region of very small $x$ and very large $A$ where the higher 
twist terms can be expected to be more important \cite{RVMcL}, 
should also be studied in more detail.

To conclude, the results of our study are encouraging. They show 
that further QCD-analysis on nuclear parton distributions is worth 
performing. For a more detailed analysis, this study can be improved 
in obvious ways:
A more quantitative error analysis should be done by performing 
a minimization of $\chi^2$ in fits to the data. In determining 
the nuclear gluon distributions by using the NMC data \cite{NMC96} 
in a more consistent manner such analysis would be required.
Also more work remains to be done in determining the nuclear ratios
for individual quark flavors. This should be done, however,
model-independently by using the measured data wherever possible. 
More systematics on mass number dependence of the structure functions  
$F_2^{\nu A}$ and $F_3^{\nu A}$ and increased statistics against 
deuterium would be helpful in extracting the nuclear ratios of 
individual parton distributions (see also \cite{KUMANOF3}). Also 
data on $F_2^A/F_2^D$ (or e.g. $F_2^A/F_2^{\rm C}$), not corrected for 
non-isoscalar effects, might be useful for constraining the difference 
between $R_u^A$  and $R_d^A$, provided that precision of the 
data is sufficient. Naturally, further measurements of the 
Drell--Yan cross sections in $pA$ collisions with more statistics 
would be very useful.

We do not expect the nuclear ratios to depend strongly on the choice 
for the (modern) parton distributions of the {\em free} proton,
but an explicit study of this is in progress. For numerical 
applications we are also preparing a package which produces scale 
dependent nuclear ratios $R_f^A(x,Q^2)$ for any parton flavour $f$ 
in an arbitrary nucleus $A$ \cite{ES98}.

Eventually, our analysis should be extended to next-to-leading 
order in the cross sections and in the scale evolution\footnote{NLO 
evolution of nuclear parton distributions is studied in \cite{KUMANO}}. 
It will also be interesting to study the role of the parton fusion 
corrections in more detail in the light of the HERA results.

\bigskip
\noindent{\bf Acknowledgements.}
We thank W. Krasny, C. Louren\c{c}o, C.A. Salgado, G. Schuler,
G. Smirnov, M. Strikman, R. Venugopalan and M. V\"anttinen for 
useful discussions, and P. McGaughey for providing us with the  
$\langle Q^2\rangle$-values of the E772 Drell--Yan data. 
We are grateful to the CFIF and GTAE centers of the Technical University 
of Lisbon for hospitality during the Hard Probe meeting in September 
1997, where part of this work was discussed and developed. 
We also thank the members of the Hard Probe Collaboration for 
helpful discussions, comments and encouragement.
This work was supported by the Academy of Finland, grant no. 27574.

\bigskip
\bigskip

\centerline{\Large \bf Appendix}
\bigskip

We use the following piecewize parametrization for the ratio
$R_{F_2}^A(x,Q_0^2)$, motivated by \cite{KJE,VARY}:

\begin{equation}
R_{F_2}(x,A) = 1 + s(x,A) + {\rm emc}(x,A) + f(x,A),
\end{equation}
with the shadowing part given by
\begin{equation}
s(x,A) = [s_0(x,A)-1]{\rm e}^{-x^2/x_0^2}\Theta(x_m-x),
\end{equation}
where $\Theta$ is a step function, and
\begin{equation}
s_0(x,A) = \frac{1+a_s k_2(1/x - 1/x_{se})p(x)}
                {1+a_s A^{p_2}(1/x - 1/x_{se})p(x)},
\end{equation}
and $p(x) = {\rm max}[1,(x/x_{se})^{p_3}]$ and
$p_3 \ge 0$.

The emc-part is controlled by a function
\begin{equation}
{\rm emc}(x,A) = (a_0+a_1x+a_2x^2+a_3x^3-1)
(1-{\rm e}^{-(x/x_{0e}^A)^2})\Theta(x_m-x),
\end{equation}
where $x_{0e}^A = x_{0e}^{\rm C}(A/12)^{p_e}$. We fix the location of 
the  minimum in $R_{F_2}(x,A)$ at $x_m=0.74$ for all nuclei. With the 
parameter $x_0$ for $s(x,A)$ and with the parameter $x_{0e}^A$ for 
emc$(x,A)$, we obtain a smoothly behaving anti-shadowing region in 
$R_{F_2}(x,A)$.

The Fermi-motion region is parametrized with
\begin{equation}
f(x,A) = \biggl[(1-f_m^A)\frac{(x-x_m)^2}{(x_f-x_m)^2} + 
f_m^A-1\biggr]\Theta(x-x_m)\\
\end{equation}
with $f_m^A = f_{m}^{\rm Ca}(A/40)^{\alpha}$,
where $\alpha$ is given by Eq.~(9) of Ref.~\cite{SLACre} with $x=x_m$.
For simplicity, we have fixed $R_{F_2}(x_f,A)=1$ for all nuclei.

The parameters $a_0, a_1, a_2, a_3$ above are determined from
the following conditions:
\begin{equation}
  e(x_e) = 1,\,\,\,\,\,\,\,\,
  e(x_m) = f_m^A,\,\,\,\,\,\,\,\,
  \frac{\partial e(x)}{\partial x}\bigg|_{x=x_{0e}} = 0,\,\,\,\,\,\,\,\,\,
  \frac{\partial e(x)}{\partial x}\bigg|_{x=x_m}    = 0,
\end{equation}
where $e(x)=a_0+a_1x+a_2x^2+a_3x^3$.
In the Table \ref{table1} below, we give the parameters
for the parametrizations of $R_{F_2}^A(x,Q_0^2)$
and $R_V^A(x<x_p,Q_0^2)$.

\begin{table}[tb]
\center
\begin{tabular}{|c|c|c|}
\hline
  parameter    & $R_{F_2}^A(x,Q_0^2)$ & $R_V^A(x<x_p,Q_0^2)$  \\
\hline
$x_{se}$         & 0.075 & 0.028\\
$x_0$            & 0.10  & 0.11\\
$x_{0e}^{\rm C}$ & 0.14  & 0.14\\
$x_e$            & 0.32  & 0.32\\
$x_m$            & 0.74  & 0.74\\
$x_f$            & 0.84  & 0.84\\
$a_s$            & 0.016 & 0.019\\
$k_2$            & 1.1   & 1.03\\
$p_2$            & 0.113 & 0.03\\
$p_3$            & 0.3   & 0.3\\
$p_e$            & 0.05  & 0.05\\
$f_m^{\rm Ca}$   & 0.87  & 0.87\\
\hline
\end{tabular}
\caption[1]{{\small The parameters used for $R_{F_2}^A(x,Q_0^2)$
and $R_V^A(x<x_p,Q_0^2)$ with $x_p = 0.09$} (see Sec.~2). }
\label{table1}
\end{table}

The form of the initial nuclear gluon ratio $R_G^A(x,Q_0^2)$ is the following:
\begin{eqnarray}
R_G^A(x,Q_0^2) &=& R_{F_2}(x,A) \biggl\{1 + N\bigg[
\exp\biggl( -\frac{(\log x-(\log x_g-\Delta_g/1.5))^2}{\Delta_g^2} \biggr) +
\nonumber\\
&&\exp\biggl( -\frac{(\log x-(\log x_g+\Delta_g/1.5))^2}{\Delta_g^2}\biggr)\bigg] \biggr\},
\label{RGQ0}
\end{eqnarray}
where $x_g$ gives the position for the additional anti-shadowing
bump in $R_G^A$, and $\Delta_g$ determines its width in $\log x$.
The amplitude $N$ for the enhancement can be directly
determined by using the momentum sum rule (\ref{MOMENTUM}).
In Figs.~\ref{RQ0}, we have $x_g= 0.09$ and $\Delta_g = 0.9$.

\end{document}